\documentclass{sig-alternate-2013}
\usepackage{subfigure}
\usepackage{color}
\usepackage{flushend}
\usepackage{times}
\usepackage[bf]{caption}
\renewcommand{\small}{\fontsize{7.5}{8.5}\selectfont}

\usepackage{stfloats}
\usepackage{paralist}
\usepackage{enumitem}

\mathchardef\Gamma="0100 \mathchardef\Delta="0101
\mathchardef\Theta="0102 \mathchardef\Lambda="0103
\mathchardef\Xi="0104 \mathchardef\Pi="0105
\mathchardef\Sigma="0106 \mathchardef\Upsilon="0107
\mathchardef\Phi="0108 \mathchardef\Psi="0109
\mathchardef\Omega="010A

\newcommand{\outline}[1]{}%

\newcommand{\descr}[1]{\smallskip\noindent\textbf{#1}}

\usepackage{xspace}
\usepackage{url}
\usepackage{graphicx}
\usepackage{latexsym}
\usepackage{amssymb}
\usepackage{amsfonts}
\usepackage{psfrag}
\usepackage{wrapfig}
\usepackage{comment}
\usepackage{color, colortbl}

\newcommand{\etal}{\frenchspacing{}\emph{et al{.}}\xspace}

\newcommand{\Comment}[1]{}

\makeatletter
\def\url@leostyle{%
  \@ifundefined{selectfont}{\def\UrlFont{}}%
  {\def\UrlFont{}}%
}
\makeatother	
\usepackage{breakurl}
\renewcommand{\footnotesize}{\fontsize{9}{10}\selectfont}

\usepackage[bookmarks=false, colorlinks=true, plainpages = false,  linkcolor = black,   citecolor = black, urlcolor = black, filecolor = black]{hyperref}

\makeatletter
\let\@copyrightspace\relax
\makeatother

\begin{document}

\title{Paying for Likes?\\Understanding Facebook Like Fraud Using Honeypots}

\numberofauthors{5}
\author{
\alignauthor{
Emiliano De Cristofaro\\
       \affaddr{University College London}\\
       \affaddr{London, U.K.}}
\and
\alignauthor{
Arik Friedman\\
       \affaddr{NICTA}\\
       \affaddr{Sydney, Australia}}
\and
\alignauthor{
Guillaume Jourjon\\
       \affaddr{NICTA}\\
       \affaddr{Sydney, Australia}}
\and
\alignauthor{
Mohamed Ali Kaafar\\
       \affaddr{NICTA \& INRIA}\\
       \affaddr{Sydney, Australia}}
\and
\alignauthor{
M. Zubair Shafiq\\
       \affaddr{The University of Iowa}\\
       \affaddr{Iowa City, IA, U.S.A.}}
}

\maketitle

\begin{abstract}
Facebook pages offer an easy way to reach out to a very large audience as they can easily be promoted using Facebook's advertising platform. Recently, the number of likes of a Facebook page has become a measure of its popularity and profitability, and an underground market of services boosting page likes, aka like farms, has emerged. Some reports have suggested that like farms use a network of profiles that also like other pages to elude fraud protection algorithms, however, to the best of our knowledge, there has been no systematic analysis of Facebook pages' promotion methods.

This paper presents a comparative measurement study of page likes garnered via Facebook ads and by a few like farms. We deploy a set of honeypot pages, promote them using both methods, and analyze garnered likes based on likers' demographic, temporal, and social characteristics. We highlight a few interesting findings, including that some farms seem to be operated by bots and do not really try to hide the nature of their operations, while others follow a stealthier approach, mimicking regular users' behavior.
\end{abstract}

\section{Introduction}
\label{sec: introduction}
Online Social Networks (OSNs) such as Facebook have become one of the primary outlets for businesses and enterprises to advertise and communicate with their customers.
Just in 2013, Facebook's net ad revenue amounted to \$6.7B, i.e., 5.64\% of the global market~\cite{emarketer1}.
One feature offered by Facebook is the concept of {\em pages}, which business can create, e.g., to display information about products and events. Users can {\em like} them to receive updates, post messages, or connect with other customers. Page likes also become part of the user's profile.
The number of likes for a given page is often considered a measure of its popularity: ChompOn estimates the
expected revenue from each like to be \$8, while other estimates range between \$3.60, \$136.38, and \$214.81~\cite{carter13like}.

To reach out to their potential audience, businesses can {\em promote} their Facebook page using targeted ads,
via {\em page like} ads.
Based on the advertiser's preferences, ads can be targeted to  users from a specific age or location group, or to users who have certain interests.
As per Facebook's guidelines, %
this is the only legitimate way to collect page likes~\cite{facebooklikebuy}.
However, a growing underground industry has emerged that provides paid services, aka {\em like farms}, to inflate the number of Facebook page likes.
Some recent press articles~\cite{guardianreport,bbc,searchengineoriginal,hufflike} have  started to look into Facebook page promotion methods and speculated that like farms use fake profiles trying to imitate real users' behavior. As these likes -- which we call {\em fake likes} -- do not correspond to a genuine interest in the advertised page, they are less valuable to businesses in terms of potential customer engagement and revenue.
Other reports~\cite{dangerousminds,veritasiumfbfraud,likeorlie} have suggested that promoting pages using legitimate Facebook ad campaigns may also garner significant amounts of fake likes. One possible explanation is that fake profiles attempt to diversify their liking activities to avoid Facebook's fraud detection algorithms. To do so, they need to click on ads and like pages other than those they are paid for.
However, to the best of our knowledge, there has been no systematic analysis
of Facebook pages' promotion methods, even though the understanding
of fake likes is arguably crucial to improve algorithms for fraud detection/mitigation in OSNs.

In this paper, we start addressing this gap with a comparative measurement study of Facebook likes garnered by means of
legitimate Facebook page like ads and by using a few underground like farms.
We set up thirteen Facebook {\em honeypot} pages and promote them using both methods.
We monitor the likes garnered by these pages, collect information about the likers
(e.g., gender, age, location, friend list, etc.), and perform a comparative analysis based on demographic, temporal, and social characteristics of the likers.

Our study highlights a few interesting findings. When targeting Facebook users worldwide, we obtain likes from only a few countries. Likers' profiles also seem to be skewed toward male profiles.
We found evidence that different like farms (with different pricing schemes) garner likes from a similar set of users and may be managed by the same operator.
We also  identified two main {\em modi operandi} of the like farms. Our results suggest that a first set of farms is operated by bots and do not really try to hide the nature of their operations,
delivering likes in bursts and forming disconnected social sub-graphs.
Other farms follow a stealthier approach, mimicking regular users' behavior, and rely on a large and well-connected network structure to gradually deliver likes while keeping a small count of likes per user.
The first strategy reflects a ``quick and dirty'' approach where likes from disposable fake
users (as also indicated by the number of terminated accounts) are delivered rapidly, as opposed to the second one,  which exhibits a stealthier approach that leverages the underlying social graph, where real users (or well-masked fake users) trickle their likes.

We did not find direct evidence that the likes garnered by the Facebook campaigns also originate from like farms. However, when comparing profiles attracted by the Facebook campaigns to those associated with like farms, we
did identify a noticeable overlap in the pages they liked overall. We also observed that likers from
Facebook campaigns liked a lot more pages than typical Facebook users, and much
closer to that observed for like farm users.

\section{Related Work}
\label{sec: related work}
Prior work has studied and detected sybil and/or fake OSN accounts by relying on tightly-knit community structures~\cite{cao12fakeosn,danezis09SybilInfer,yang11socialnetworksybils,yu08sybillimit,yu06sybilguard}.
Findings revealed by our work also highlight several characteristics about the social structure and activity of fake profiles attracted by the honeypot pages,
e.g., their interconnected nature or the activity bursts. In fact,
our analysis does not only confirm a few insights used by sybil detection algorithms but also reveals new patterns
that could complement them.
A few {\em passive} measurement studies have also focused on characterizing fake user accounts and their activity.
Nazir \etal~\cite{nazir10facebookphantomprofiles} studied phantom profiles in Facebook gaming applications, while
Thomas \etal~\cite{thomas11suspendedaccounts} analyzed over 1.1 million accounts suspended by Twitter.
Gao \etal~\cite{gao10socialspamcampaigns} studied spam campaigns on Facebook originating from approximately 57,000 user accounts.
Yang \etal~\cite{yang12spammersocialnetwork} performed an empirical analysis of social relationships between spam accounts on Twitter, and Dave \etal~\cite{dave2012measuring} proposed a methodology to measure and fingerprint click-spam in ad networks.
Our work differs from these studies as they all conduct passive measurements, whereas we rely on the deployment of several honeypot pages and (paid) campaigns to actively engage with fake profiles.

Stringhini \etal~\cite{stringhini10spammerssocialnetworks} and Lee \etal~\cite{lee10socialspamhoneypots} created honeypot profiles in Facebook, MySpace, and Twitter to detect spammers. Their work differs from ours in that
(1) their honeypot profiles were designed to look legitimate, while our honeypot pages explicitly indicated they were not ``real'' (to deflect real profiles), and (2) our honeypot pages {\em actively} attracted fake profiles by means of paid campaigns, as opposed to passive honeypot profiles.
Also, Thomas \etal \cite{thomas13traffickingfraudtwitteraccounts} analyzed trafficking of fake accounts in Twitter.
They bought fake profiles from 27 merchants and developed a classifier to detect these fake accounts.
In a similar study, Stringhini \etal~\cite{stringhini13twitterfollower} analyzed the market of {\em Twitter followers}, which, akin to Facebook like farms, provide Twitter followers for sale.
Note that Twitter follower markets differ from Facebook like farms as Twitter entails a {\em follower-followee} relationship among users, while Facebook friendships imply a bidirectional relationships. Also, there is no equivalent of liking a Facebook page in the Twitter ecosystem.

Beutel \etal~\cite{beutel2013copycatch} proposed a technique to detect fake likes based on identifying groups of users who liked a set of pages within a given time period. However, their technique does not rely on ground truth data, so it is unclear whether or not the detection mechanism blocks all fake likes, or actually only those exhibiting a certain pattern.
By contrast, we focus on actively measuring like fraud activities by means of honeypots, i.e., attracting fake likes to empty pages, through payment. We elicit and study ground truth data, and highlight how some like farms actually try to emulate behavior of regular users and thereby stay below the detection radar. Nonetheless, our work serves as the starting point for improved fake like detection and can complement techniques from~\cite{beutel2013copycatch}.

Finally, a few investigative press reports~\cite{guardianreport,bbc,searchengineoriginal,veritasiumfbfraud} have also looked into Facebook page ads and underground like farms, however, without any systematic analysis of Facebook pages' promotion methods.

\begin{table*}[t!]
\small
\centering
\begin{tabular}{|l|l|l|l|r|r|r|r|r|}
\hline
  \bf{Campaign ID} & \bf{Provider} & \bf{Description} & \bf{Location} & \bf{Budget} & \bf{Duration} & \bf{Monitoring} &
  \bf{\#Likes} & \bf{\#Terminated}\\
\hline
FB-USA & Facebook.com & Page like ads & USA & \$6/day  & 15 days & 22 days & 32 & 0\\
FB-FRA & Facebook.com & Page like ads & France & \$6/day & 15 days &22 days & 44 & 0 \\
FB-IND & Facebook.com & Page like ads & India & \$6/day  & 15 days &22 days & 518 & 2 \\
FB-EGY & Facebook.com & Page like ads & Egypt & \$6/day & 15 days &22 days & 691 & 6\\
FB-ALL & Facebook.com & Page like ads & Worldwide & \$6/day  & 15 days &22 days &484 & 3\\
\hline
BL-ALL & BoostLikes.com & 1000 likes   &Worldwide & \$70.00 & 15 days &-& - & -\\
BL-USA & BoostLikes.com & 1000 likes  & USA only & \$190.00 & 15 days &22 days & 621 & 1\\
\hline
SF-ALL & SocialFormula.com & 1000 likes  & Worldwide & \$14.99 & 3 days  &10 days & 984 & 11\\
SF-USA & SocialFormula.com & 1000 likes & USA & \$69.99 & 3 days &10 days & 738 & 9\\
\hline
AL-ALL & AuthenticLikes.com & 1000 likes   & Worldwide & \$49.95 & 3-5 days & 12 days &755 & 8 \\
AL-USA & AuthenticLikes.com & 1000 likes  & USA & \$59.95& 3-5 days &22 days & 1038 & 36 \\
\hline
MS-ALL & MammothSocials.com & 1000 likes & Worldwide & \$20.00 & -  &-  & - & -\\
MS-USA & MammothSocials.com & 1000 likes & USA only & \$95.00 & - &12 days & 317 & 9\\
\hline
\end{tabular}
\caption{Facebook and like farm campaigns used to promote our Facebook honeypot pages.}
\label{tbl:measurements}
\end{table*}

\section{Methodology}
\label{sec: data}

This section presents the methodology used to deploy and monitor Facebook honeypot pages and to promote them using both Facebook page like ads and like farms.

\descr{Honeypot Pages.} We created 13 Facebook pages called ``Virtual Electricity'' and intentionally kept them empty (i.e., no posts or pictures). Their description included: {\em ``This is not a real page, so please do not like it.''}
5 pages were promoted using legitimate Facebook (FB) ad campaigns targeting users, respectively, in USA, France, India, Egypt, and worldwide.
The remaining 8 pages were promoted using 4 popular like farms BoostLikes.com (BL), SocialFormula.com (SF), AuthenticLikes.com (AL), and MammothSocials.com (MS), targeting worldwide or USA users.

In Table \ref{tbl:measurements}, we provide details of the honeypot pages, along with the
corresponding ad campaigns. All campaigns were launched on March 12, 2014, using a different administrator account (owner) for each page. Each Facebook campaign was budgeted at a maximum of \$6/day to a total of \$90 for 15 days. The price for buying likes varied across likes farms: BoostLikes charged the highest price for ``100\% real likes'' (\$70 and \$190 for 1000 likes in 15 days from, respectively, worldwide and USA).
Other like farms also claimed to deliver likes from ``genuine'', ``real'', and ``active'' profiles, but promised to deliver them in fewer days.
Overall, the price of 1000 likes varied between \$14.99--\$70 for worldwide users and \$59.95--\$190 for USA users.

\descr{Data Collection.} We monitored the ``liking'' activity on the honeypot pages by crawling them, using Selenium web driver~\cite{selenium}, every 2 hours to check for new likes. %
At the end of the campaigns, we reduced the monitoring frequency to once a day, and stopped monitoring when a page did not receive a like for more than a week.
We used Facebook's reports tool for page administrators, which provides a variety of aggregated statistics
about attributes and profiles of page likers. Facebook also provides these statistics for the global Facebook population. Since a majority of Facebook users do not set the visibility of their age and location to public~\cite{Chaabane2012}, we used these reports to collect statistics about likers' gender, age, country, home and current town.\footnote{As stated in \cite{FB_Ads_Optimisation}, Facebook uses public {\em and} private attributes to provide aggregated statistics about users who clicked on page like ads, e.g., current location is determined based on IP address.}  Later, in Section \ref{sec: measurement}, we will use these statistics to compare distributions of our honeypot pages' likers to that of the overall Facebook population.
We also crawled public information from the likers' profiles, obtaining the lists of liked pages as well as friend lists, which are not provided in the reports. Overall, we identified more than 6.3 millions total likes by users who liked our honeypot pages and more than 1 million friendship relations. %

We acknowledge that our limited budget allows us to only monitor 13 honeypots for a few weeks. Note, however, that our work is a first-of-a-kind, exploratory study of like fraud practices -- as it will become clear in the rest of the paper, our methodology actually allows us to derive several interest findings (which could be further explored, in future work, with larger/more diverse campaigns).

\descr{Campaign Summary.} In Table~\ref{tbl:measurements}, we report the total number of likes garnered by each campaign, along with the number of days we monitored the honeypot pages. Note that the BL-ALL and MS-ALL campaigns remained inactive, i.e., they did not result in any likes even though we were charged in advance.
We tried to reach the like farm admins several times but received no response.
Overall, we collected a total of 6,292 likes (4,523 from like farms and 1,769 from Facebook ads).
The largest number of likes were garnered by AL-USA, the lowest (excluding inactive campaigns) by FB-USA.

\descr{Ethics Considerations.} Although we only collected openly available data,
we did collect (public) profile information from our honeypot pages' likers, e.g., friend lists and page likes. We could not request consent but enforced a few mechanisms to protect user privacy: all data were encrypted at rest and not re-distributed, and no personal information was extracted, i.e., we only analyzed aggregated statistics.
We are also aware that paying farms to generate fake likes might raise ethical concerns, however, this was crucial to create the honeypots and observe the like farms' behavior. We believe that the study will help, in turn, to understand and counter these activities. Also note that the amount of money each farm received was small (\$190 at most) and that this research was reviewed and approved by the NICTA legal team.

\begin{figure}[t!]
\centering
\includegraphics[width=0.95\columnwidth]{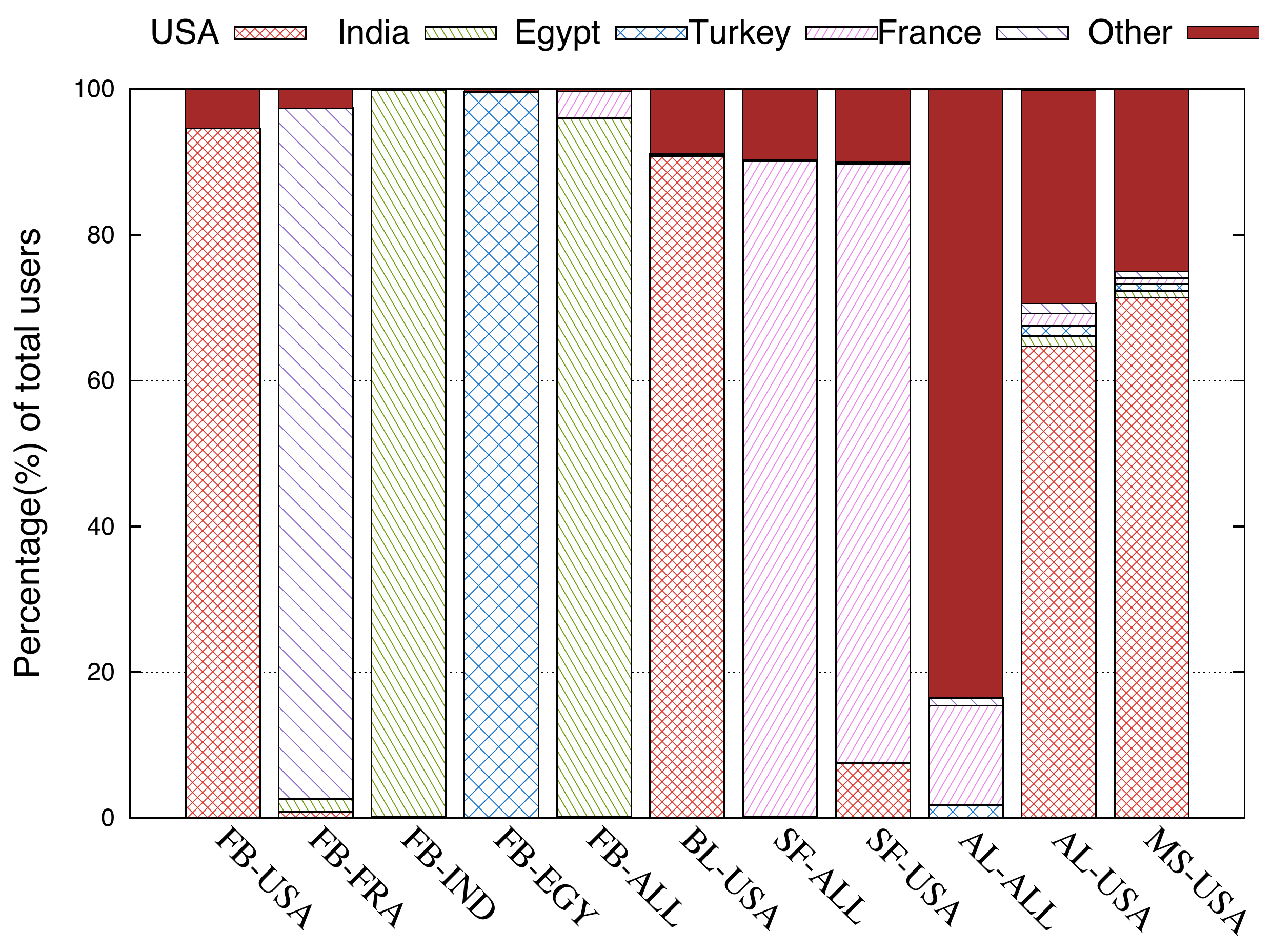}
\vspace{-0.15cm}
\caption{Geolocation of the likers (per campaign). }
\label{fig:geolocation}
\end{figure}

\section{Analysis}
\label{sec: measurement}
We now compare the characteristics of the likes garnered by the honeypot pages promoted via legitimate Facebook campaigns and those obtained via like farms.

\subsection{Location and Demographics Analysis}
\label{subsec:demographics}

\descr{Location.} For each campaign, we looked at the distribution of likers' countries: as shown in Figure~\ref{fig:geolocation}, for the first four Facebook campaigns (FB-USA, FB-FRA, FB-IND, FB-EGY), we mainly received likes from the targeted country (87--99.8\%), even though FB-USA and FB-FRA generated
a number of likes much smaller than any other campaign.
When we targeted Facebook users worldwide (FB-ALL), we almost exclusively received likes from India (96\%).
Looking at the like farms, most likers from SocialFormula were based in Turkey, regardless of whether we requested a US-only campaign. The other three farms delivered likes complying to our requests, e.g., for US-only campaigns, the pages received a majority of likes from US profiles.

\descr{Other Demographics.}
In Table~\ref{tbl:attributes}, we show the distribution of likers' gender and age, and also compare them to the global Facebook network (last row). The last column reports the KL-divergence between the age distribution of the campaign users and that of the entire Facebook population, highlighting large divergence for FB-IND, FB-EGY, and FB-ALL, which are biased toward
younger users. These three campaigns also appear to be skewed toward male profiles.
In contrast, the demographics of likers from SocialFormula and, to a lesser extent, AuhtenticLikes and MammothSocials, are much more similar to those of the entire network, even though male users are still over-represented. %

\begin{table}[t!]
\setlength{\tabcolsep}{3pt}
\small
\centering
\begin{tabular}{|l|c|cccccc|c|}
\hline
 {\bf{Campaign}} &  {\bf Gender } & \multicolumn{6}{c}{\bf{Age Distribution (\%)}} & \\
 {\bf{ID}} & {\bf \% F/M} &{\bf 13-17} & {\bf 18-24}  & {\bf 25-34} & {\bf 35-44} & {\bf 45-54}  & \multicolumn{1}{c}{\bf 55+} & {\bf KL}  \\ [0.3ex]
\hline
 FB-USA& 54/46 & 54.0&	27.0&	6.8&	6.8&	1.4&	4.1 & 0.45 \\[0.3ex]
 FB-FR	&46/54&60.8&	20.8&	8.7&	2.6&	5.2&	1.7 & 0.54 \\[0.3ex]
FB-IND	&\bf{7/93}&52.7&	43.5&	2.3&	0.7&	0.5&	0.3 & \bf{1.12} \\[0.3ex]
FB-EGY	&\bf{18/82}&54.6&	34.4&	6.4&	2.9&	0.8&	0.8 & \bf{0.64} \\[0.3ex]
FB-ALL	&\bf{6/94}&51.3&	44.4&	2.1&	1.1&	0.5&	0.6 & \bf{1.04}\\[0.3ex] \hline
BL-USA	&53/47&34.2&	54.5&	8.8&	1.5&	0.7&	0.5 & 0.60\\[0.3ex] \hline
SF-ALL	&37/63&19.8&	33.3&	21.0&	15.2&	7.2&	2.8 & 0.04\\[0.3ex]
SF-USA	&37/63&22.3	&34.6&22.9&	11.6&	5.4&	2.9 & 0.04 \\[0.3ex] \hline
AL-ALL	&42/58&15.8&	52.8&	13.4&	9.7&	5.2&	3.0 & 0.12 \\[0.3ex]
AL-USA	&31/68&7.2&	41.0&	35.0&	10.0&	3.5&	2.8 & 0.09\\[0.3ex] \hline
MS-USA	&26/74&8.6&	46.9&	34.5&	6.4&	1.9&	1.4 & 0.17\\[0.3ex] \hline \hline
Facebook	&46/54&14.9&	32.3&	26.6&	13.2&	7.2&	5.9 & -- \\ \hline
\end{tabular}
\caption{Gender and age statistics of likers.}%
\label{tbl:attributes}
\end{table}

\begin{figure*}[ht!]
\centering
	\subfigure[\small Facebook Campaigns]{\includegraphics[width=0.925\columnwidth]{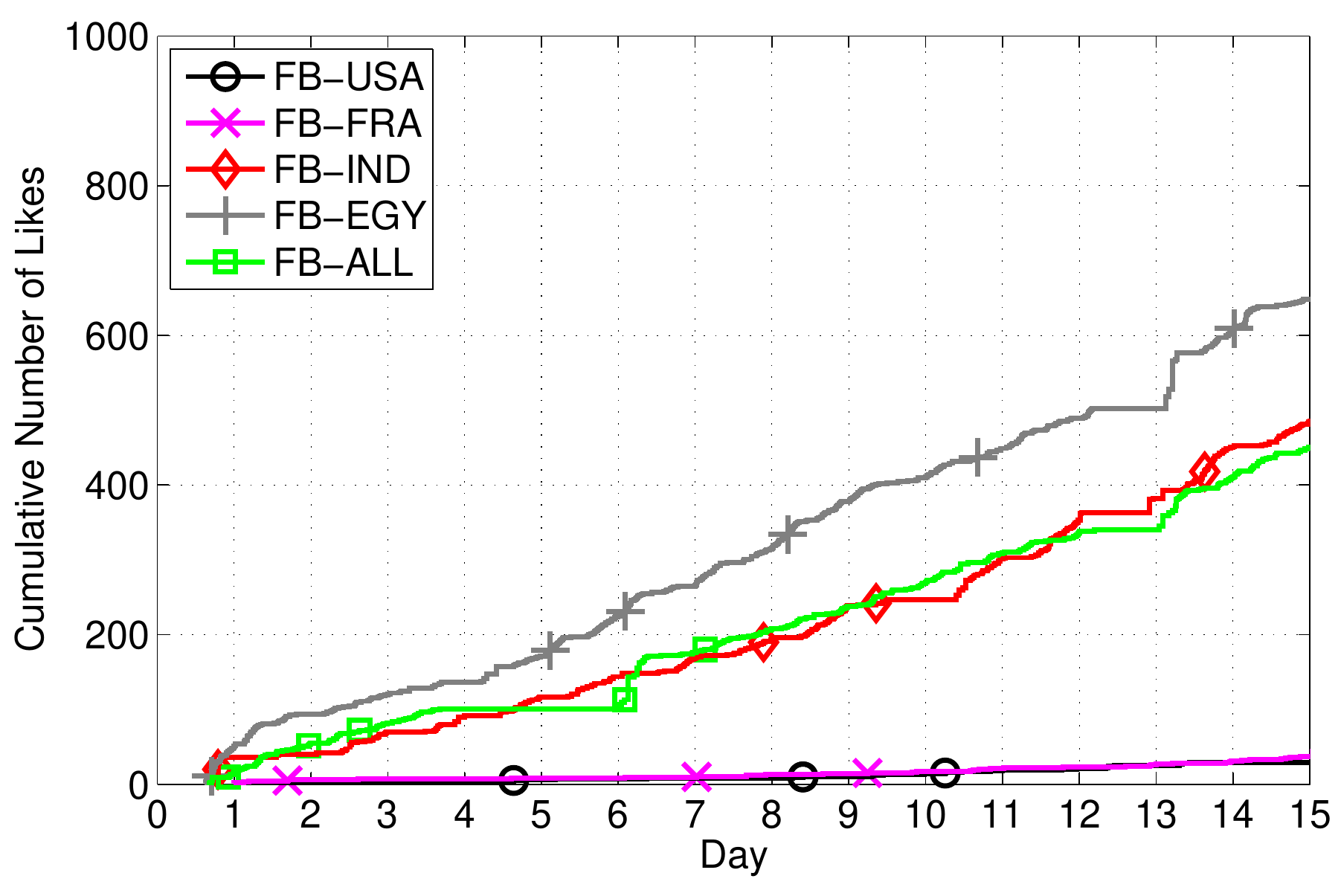}
         \label{likes_timeseries_facebook}}
         \subfigure[\small Like Farms]{\includegraphics[width=0.925\columnwidth]{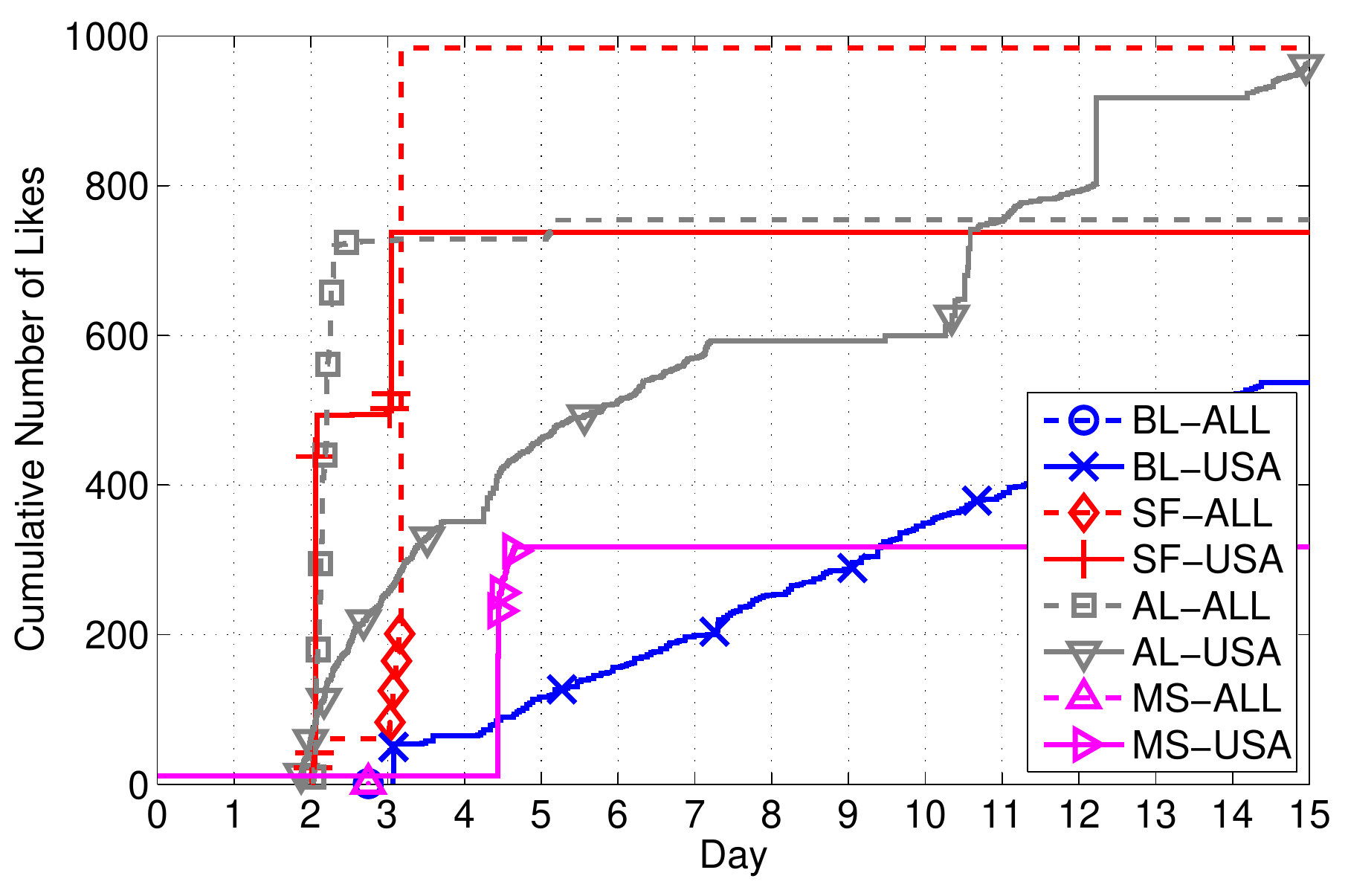}
         \label{likes_timeseries_underground}}
\vspace{-0.15cm}
\caption{Time series of cumulative number of likes for Facebook and like farms campaigns.}
\label{Fig:results:temporalAccumulcation}
\end{figure*}

\subsection{Temporal Analysis}
We also analyzed temporal patterns observed for each of the campaigns. In Figure~\ref{Fig:results:temporalAccumulcation}, we plot the cumulative number of likes observed on each honeypot page over our observation period (15 days). We observe from Figure~\ref{likes_timeseries_underground} that all the like farm campaigns, except BoostLikes, exhibit a very similar trend with a few bursts of a large number of likes. Specifically, for the SocialFormula, AuthenticLikes, and MammothSocials campaigns, likes were garnered within a short period of time of two hours. With AuthenticLikes, we observed likes from more than 700 profiles within the first 4 hours of the second day of data collection. Interestingly, no more likes were observed later on.
On the contrary, the BoostLikes campaign targeting US users shows a different temporal behavior: the trend is actually comparable to that observed in the Facebook Ads campaigns (see Figure~\ref{likes_timeseries_facebook}). The number of likes steadily increases during the observation period and no abrupt changes are observed.

This suggests that two different strategies may be adopted by like farms. On the one hand, the abrupt increase in the cumulative number of likes happening during a short period of time might likely be due to %
automated scripts operating a set of fake profiles. These profiles are instrumented to satisfy the number of likes as per the customer's request.
On the other hand, BoostLikes's strategy, which resembles the temporal evolution in Facebook campaigns,  seems to rely on the underlying social graph, possibly constituted by fake profiles operated by humans. Results presented in the next section corroborate the existence of these two strategies.

\begin{table*}[ht!]
\small
\centering
\begin{tabular}{|c|c|c|c|c|c|c|}
\hline
\bf{Provider} & \bf{\# Likers} & \bf{\# Likers with} & \bf{Avg ($\pm$ Std)} & {\bf Median} & \bf{\# Friendships}  & \bf{\# 2-Hop Friendship}  \\
 &  & \bf{Public Friend Lists}  &  \bf{\#Friends} & {\bf \#Friends} & \bf{Between Likers} & \bf{Relations Between Likers}\\ \hline
Facebook.com & 1448 &  261 (18.0\%) &  315 $\pm$ 454 & 198 &  6 & 169\\
BoostLikes.com & 621 &  161 (25.9\%) & 1171 $\pm$ 1096 &  850 & 540 & 2987  \\
SocialFormula.com & 1644 & 954 (58.0\%) &  246 $\pm$ 330& 155 &50 & 1132  \\
AuthenticLikes.com & 1597 &  680 (42.6\%) & 719 $\pm$ 973& 343 & 64 & 1174  \\
MammothSocials.com & 121 & 62 (51.2\%) & 250 $\pm$ 585& 68 & 4 & 129 \\
ALMS & 213 & 101 (47.4\%) &  426 $\pm$ 961 & 46 &  27 & 229  \\
\hline
\end{tabular}
\caption{Likers and friendships between likers.}
\label{tab:LikersAndFriendships}
\end{table*}

\subsection{Social Graph Analysis}
Next, we evaluated the social graph induced by the likers' profiles.
To this end, we associated each user with one of the like farm services based on the page they liked. Note that a
few users liked pages in multiple campaigns, as we will discuss in Section~\ref{sec:likeAnalysis}.
A significant fraction of users actually liked pages corresponding to both the AuthenticLikes and the MammothSocials campaigns (see Figure~\ref{fig:like-distribution}): we put these users into a separate group, labelled as ALMS. Table~\ref{tab:LikersAndFriendships} summarizes the number of likers associated with each service, as well as additional details about their friendship networks. Note that the number of likers reported for each campaign in Table~\ref{tab:LikersAndFriendships} is different from the number of campaign likes (Table~\ref{tbl:measurements}), since some users liked more than one page.

Many likers kept their friend lists private: this occurred for almost 80\% of likers in the Facebook campaigns, about 75\% in the BoostLikes campaign, and much less frequently for the other like farm campaigns ($\sim$40--60\%). The number and percentage of users  with public friend lists are reported in Table~\ref{tab:LikersAndFriendships}.
The fourth column reports the average number of friends ($\pm$ the standard deviation) for profiles with visible friend lists, and the fifth column reports the median.
Some friendship relations may be hidden, e.g., if a friend chose to be invisible in friend lists, thus, these numbers only represent a {\em lower bound}. The average number of friends of users associated with the BoostLikes campaign (and to a smaller extent, the AuthenticLikes campaign) was much higher than the average number of friends observed elsewhere.
\begin{figure*}[ht!]
\centering
         \subfigure[Direct friendship relations\label{fig:friendship}]{\includegraphics[width=0.8\columnwidth]{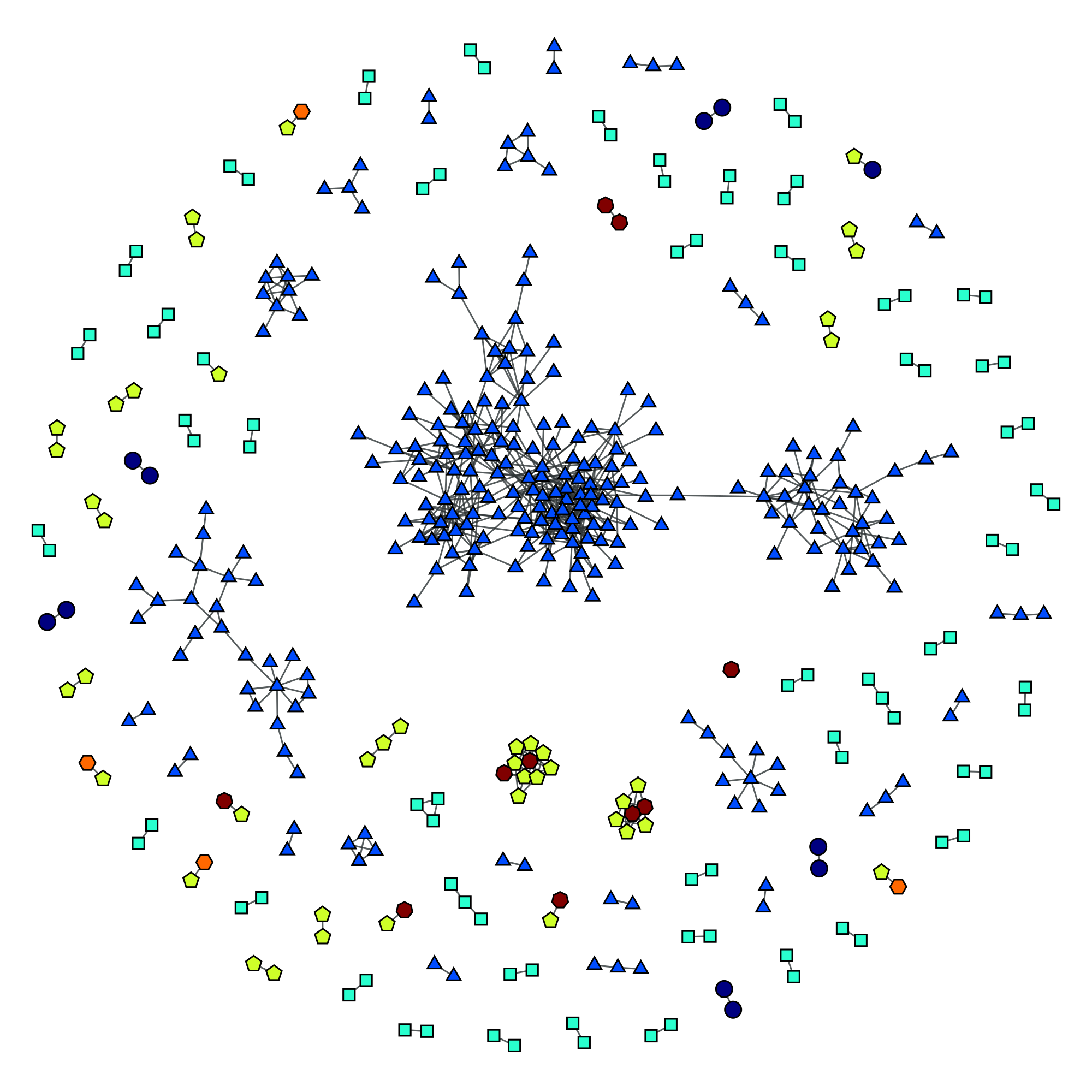}}
         \subfigure{\includegraphics[width=0.35\columnwidth]{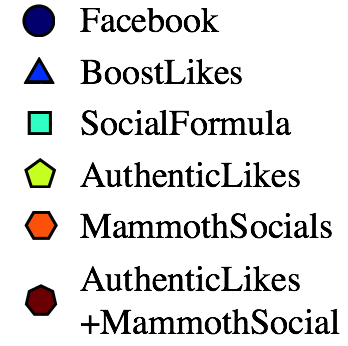}}
         \addtocounter{subfigure}{-1}
         \subfigure[2-hop friendship relations\label{fig:2hop_friendship}]{\includegraphics[width=0.8\columnwidth]{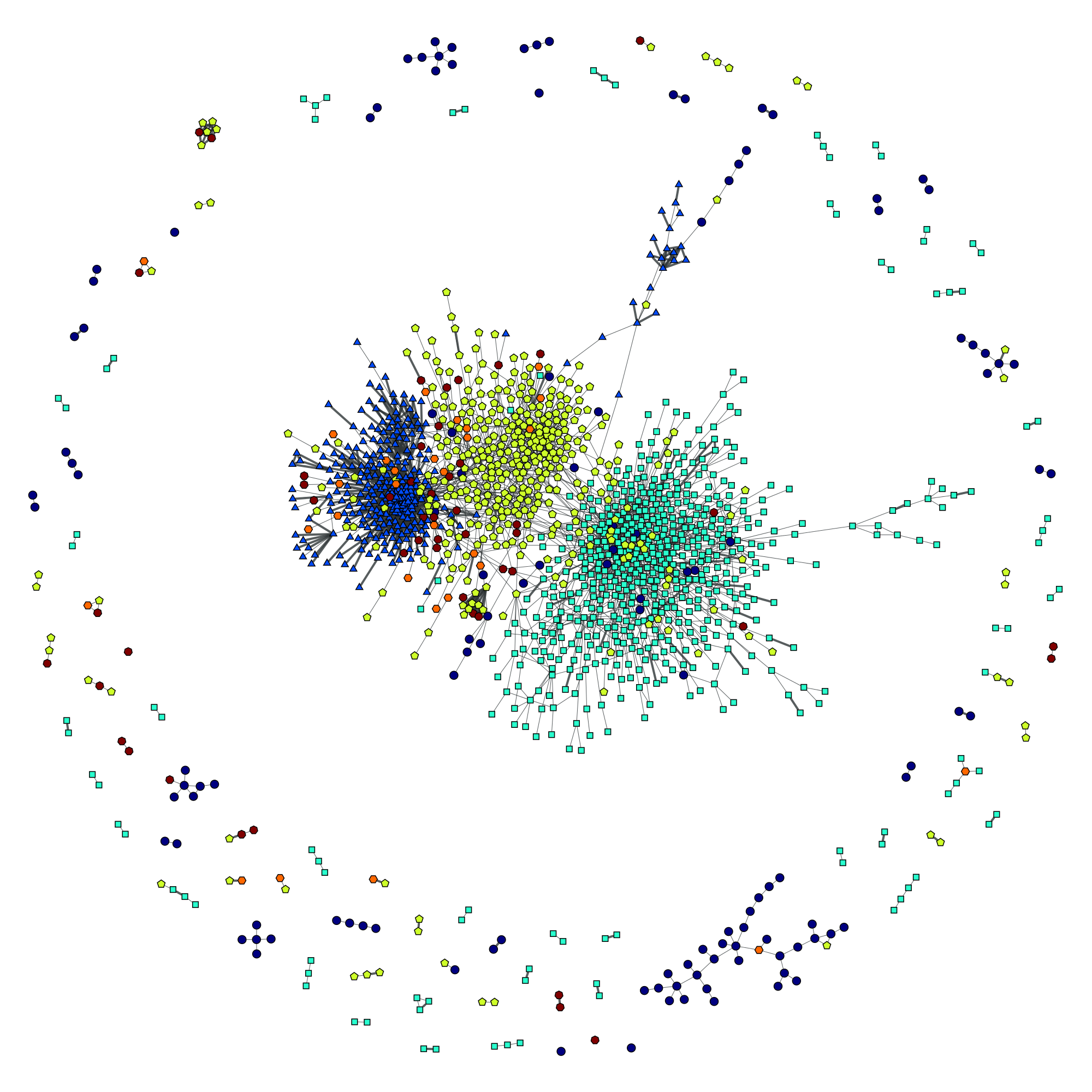}}
\caption{Friendship relations between likers of different campaigns.}
\label{fig:friendshipGraph_withALMS}
\end{figure*}

To evaluate the social ties between likers, we looked at friendship relations between likers (either originating from the same campaign provider or not), ignoring friendship relations with Facebook users who did not like any of our pages. Table~\ref{tab:LikersAndFriendships} (sixth column) reports, for each provider, the overall number of friendship relationships between likers that involved users associated with the provider.

In Figure~\ref{fig:friendship}, we plot the social graph induced by such friendship relations (likers who did not have friendship relations with any other likers were excluded from the graph). %
Based on the resulting social structure, we suggest that: %
\begin{enumerate}
\item
 Dense relations between likers from BoostLikes point to an interconnected network of real users, or fake users who mimic complex ties to pose as real users;
\item
 The pairs (and occasionally triplets) that characterize SocialFormula likers might indicate a different strategy of constructing fake networks, mitigating the risk that identification of a user as fake would consequently bring down the whole connected network of fake users; and
\item
 The friendship relations between AuthenticLikes and MammothSocials likers might indicate that the same operator manages both services.
\end{enumerate}

We also considered indirect links between likers, through mutual friends. Table~\ref{tab:LikersAndFriendships} reports the overall number of 2-hop relationships between likers from the associated provider. Figure~\ref{fig:2hop_friendship} plots the relations between likers who either have a direct  relation or a mutual friend, clearly pointing to the presence of relations between likers from the same provider. These tight connections, along with the number of their friends, suggest that we only see a small part of these networks. (In fact, like farms sell packages of as many as 50k likes.) For SocialFormula, AuthenticLikes, and MammothSocials, we also observe many isolated pairs and triplets of likers who are not connected. One possible explanation is that farm users create fake Facebook accounts and keep them separate from their personal accounts and friends. In contrast, the BoostLikes network is well-connected.

\subsection{Page Like Analysis}
\label{sec:likeAnalysis}
\begin{figure*}[ht!]
\centering
         \subfigure[\small Facebook Campaigns]{\includegraphics[width=0.925\columnwidth]{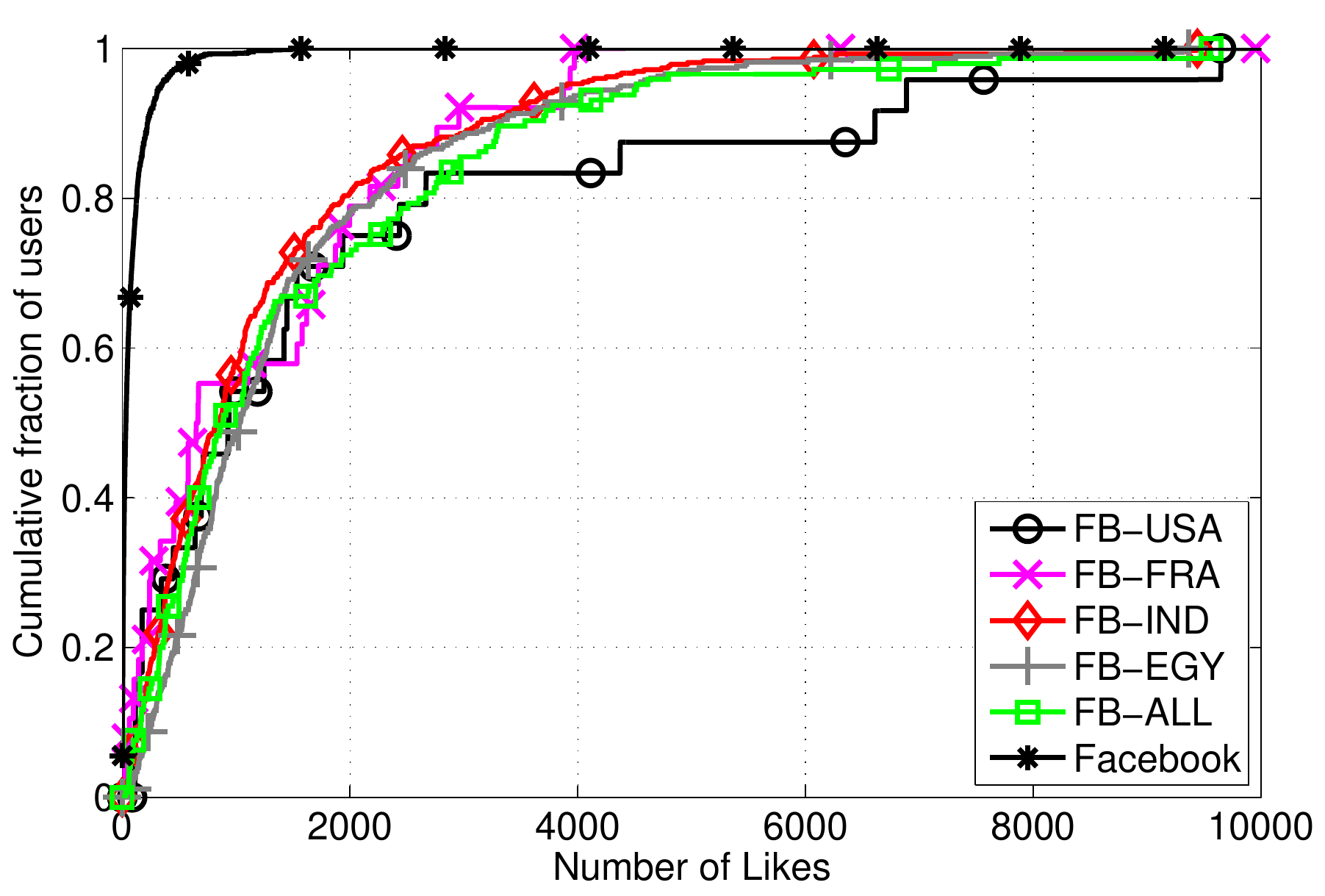}\label{fig:like-distribution-facebook}}
         \subfigure[\small Like Farms]{\includegraphics[width=0.925\columnwidth]{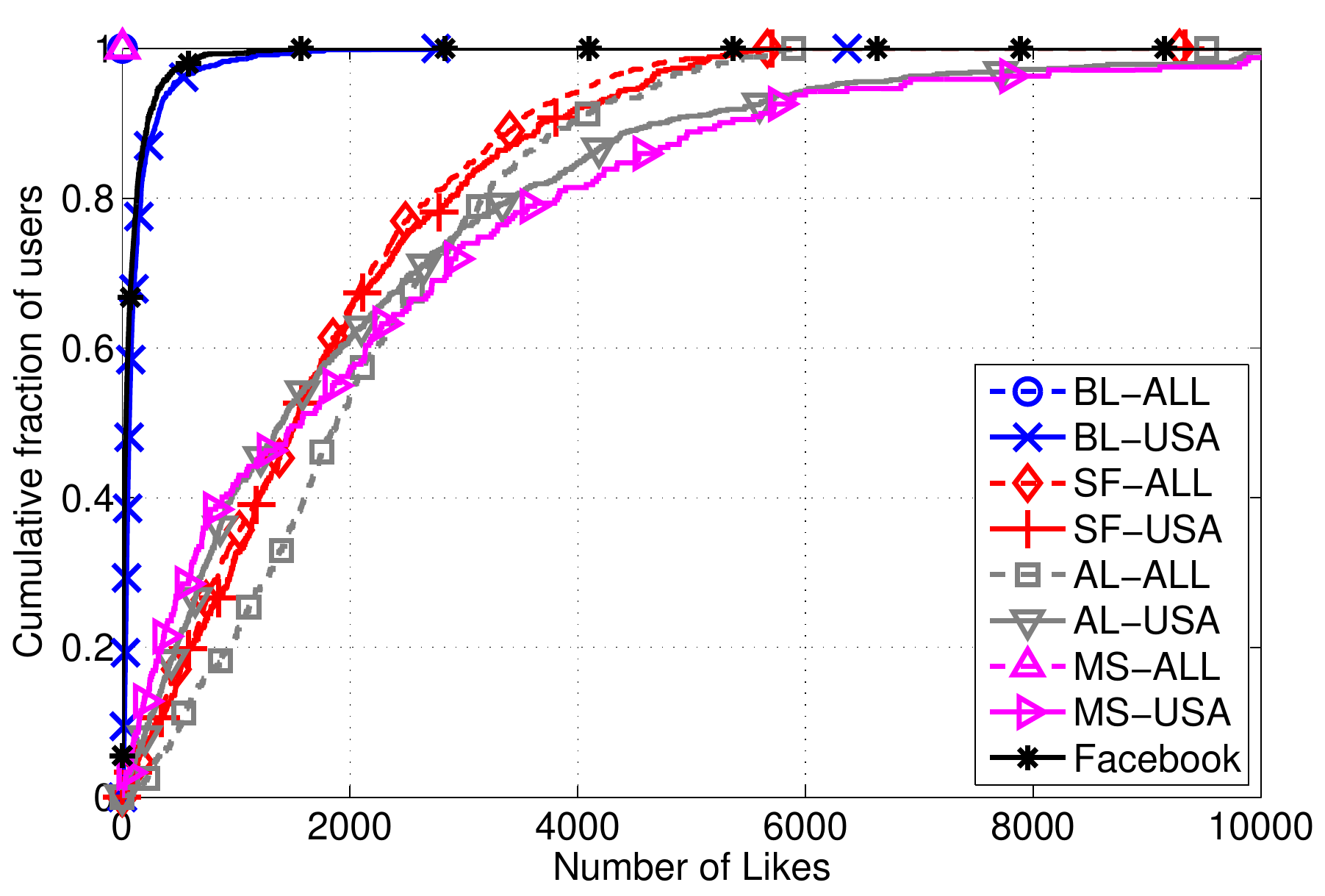}\label{fig:like-distribution-farms}}
\vspace{-0.15cm}
\caption{Distribution of the number of likes by users in Facebook and like farm campaigns.}
\label{fig:like-distribution}
\end{figure*}

We then looked at the {\em other} pages liked by profiles attracted to our honeypot pages.
In Figure \ref{fig:like-distribution-facebook} and \ref{fig:like-distribution-farms}, respectively,
we plot the distribution of the number of page likes for Facebook ads' and like farm campaigns' users.
To draw a baseline comparison, we also collected page like counts from a random set of 2000 Facebook users, extracted from an unbiased sample of Facebook user population. The original sample was crawled for another project~\cite{chen2013much}, obtained by randomly sampling Facebook public directory which lists all the IDs of searchable profiles.

We observed a large variance in the number of pages liked, ranging from 1 to 10,000.
The median page like count ranged between 600 and 1000 for users from the Facebook campaigns and between 1200 and 1800 for those from like farm campaigns, with the exception of the BL-USA campaign (median was 63).
In contrast, the median page like count for our baseline Facebook user sample was 34.
The page like counts of our baseline sample mirrored numbers reported in prior work, e.g.,
according to \cite{allfacebook}, the average number of pages liked by Facebook users amounts to roughly 40.
In other words, our honeypot pages attracted users that tend to like significantly more pages than regular Facebook users.
Since our honeypot pages both for Facebook and like farm campaigns explicitly indicated they were not ``real'', we argue that a vast majority of the garnered likes are fake.
We argue that these users like a large number of pages because they are probably reused for multiple ``jobs'' and also like ``normal'' pages to mimic real users.\footnote{Facebook does not impose any limit on the maximum number of page likes per user.}

\begin{figure*}[t!]
\centering
         \subfigure[\small Page Like]{\includegraphics[width=0.925\columnwidth]{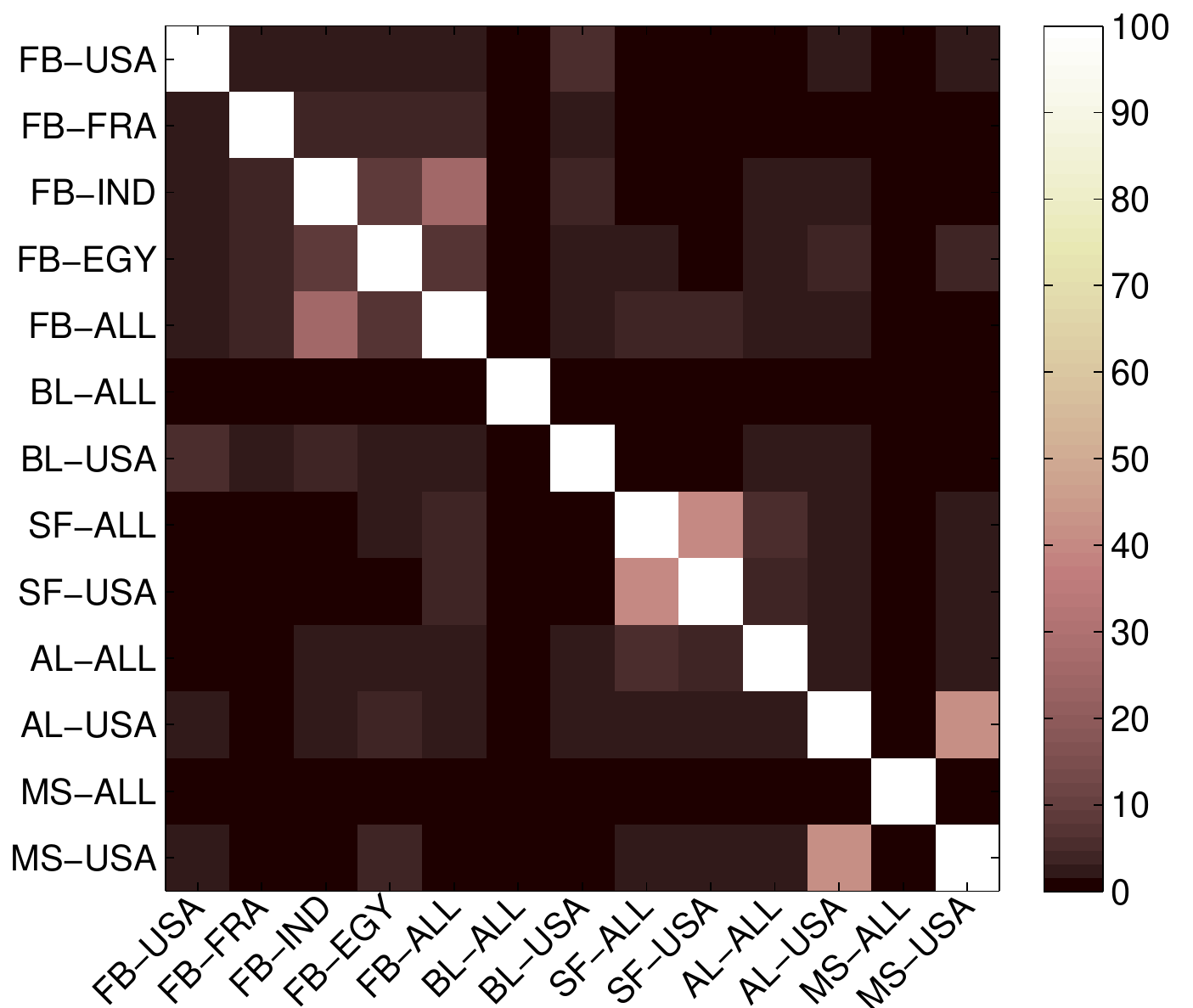}\label{fig:similarity-matrix-a}}
         \subfigure[\small User]{\includegraphics[width=0.925\columnwidth]{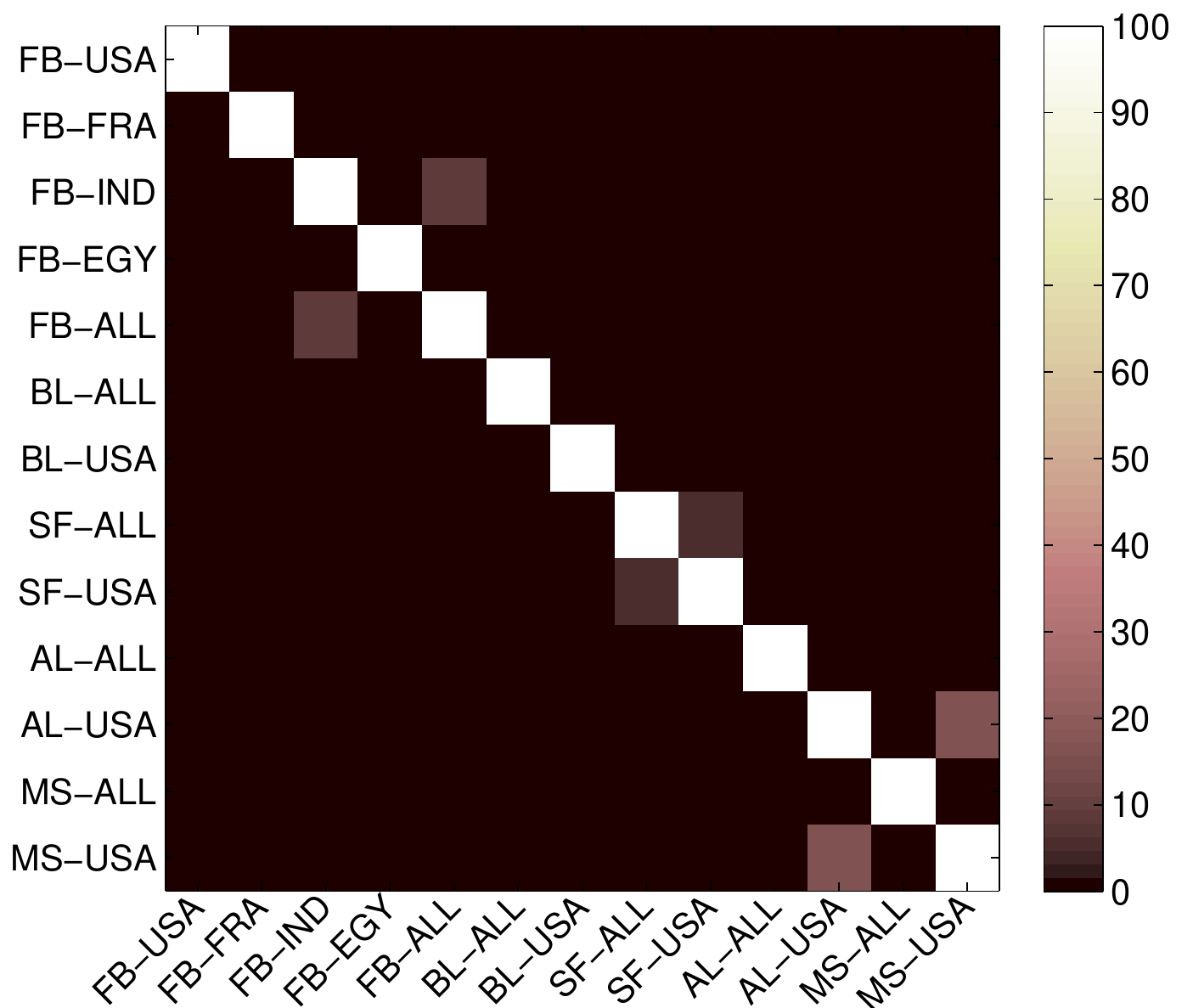}
\label{fig:similarity-matrix-b}}
\vspace{-0.15cm}
\caption{Jaccard index similarity ($\times 100$) matrices of page likes and likers across different campaigns}
\label{fig:similarity-matrix}
\end{figure*}

To confirm our hypothesis, for each pair of campaigns, we plot their Jaccard similarity.
Specifically, let $S_k$ denote the set of pages liked by a user $k$: the Jaccard similarity between the set of likes by likers of two campaigns $A$ and $B$, which we plot in Figure~ \ref{fig:similarity-matrix-a}, is defined as $|A \cap B|/| A \cup B|$,
where $A = \bigcup_{\forall i \in A} S_i $ and $ B = \bigcup_{\forall j \in B} S_j $. %
We also plot, in Figure~\ref{fig:similarity-matrix-b}, the similarity between
$A' = \bigcup_{\forall i \in A} i $ and $ B' = \bigcup_{\forall j \in B} j $, i.e., the similarity between the set of likers of the different campaigns.

Note from Figure \ref{fig:similarity-matrix} that FB-IND, FB-EGY, and FB-ALL have relatively large (Jaccard) similarity with each other.
In addition, the SF-USA and SF-ALL pair and the AL-USA and MS-USA pair also have relatively large Jaccard similarity.
These findings suggest that the same fake profiles are used in multiple campaigns by a like farm (e.g., SF-ALL and SF-USA). Moreover, some fake profiles seem to be shared by different like farms (e.g., AL-USA and MS-USA), suggesting that they are run by the same operator.

\section{Concluding Remarks}
\label{sec: conclusions}
This paper presented a comparative measurement study of Facebook page promotion methods, aiming to shed light on like farms' activities.
We identified two main {\em modi operandi}: (1) some farms, like SocialFormula and AuthenticLikes, seem to be operated by bots and do not really try to hide the nature of their operations, as demonstrated by large bursts of likes and the limited number of friends per profile; (2) other farms, like BoostLikes, follow a much stealthier approach, aiming to mimic regular users' behavior, and rely on their large and well-connected network structure to disseminate the target likes while keeping a small count of likes per user. For the latter, we also observed a high number of friends per profile and a ``reasonable'' number of likes.

A month after the campaigns, we checked whether or not likers' accounts were still active:
as shown in Table~\ref{tbl:measurements}, only one account associated with BoostLikes was terminated, as opposed to 9, 20, and 44 for the other like farms. 11 accounts from the regular Facebook campaigns were also terminated.
Although occurring not so frequently, the accounts' termination might be indicative of the disposable nature of fake accounts on most like farms, where ``bot-like" patterns are actually easy to detect.
It also mirrors the challenge Facebook is confronted by, with like farms such as BoostLikes that exhibit patterns closely resembling real users' behavior, thus making fake like detection quite difficult.

We stress that our findings do not necessarily imply that advertising on Facebook is ineffective, since our campaigns were specifically designed to avert real users. However, our work provides strong evidence that likers attracted on our honeypot pages, even when using legitimate Facebook campaigns, are significantly different from typical Facebook users, which confirms the concerns about the genuineness of these likes.
We also show that most fake likes exhibit some peculiar characteristics -- including demographics, likes, temporal and social graph patterns -- that can and should be exploited by like fraud detection algorithms.

Besides the design of detection techniques, items for future work include larger and more diverse honeypots measurements as well as longer observation of removed likes. Also, as suggested in prior work~\cite{beutel2013copycatch},  fake likes might be generated via fake accounts, malware, malicious browser extensions, and social engineering, thus prompting the need for further investigation of fake likes' origin.

\descr{Acknowledgments.} We would like to thank the reviewers and our shepherd for their feedback and useful comments. We are also grateful to Gianluca Stringhini for reviewing a draft of the paper.

%
%\bibliographystyle{abbrv}
%\bibliography{sigproc}

\end{document}